\documentstyle[psfig]{mn} 
\newcommand{\ea}{{\it et~al.\/} }
\newcommand{\dg}{\nobreak^\circ}

\newcommand{\ho}{\mbox{$\mbox{H}_0$} }
\newcommand{\simlt}{\mbox{$\stackrel{<}{_{\sim}}$} }
\newcommand{\simgt}{\mbox{$\stackrel{>}{_{\sim}}$} }

\newcommand{\kmspmpc}{\,\hbox{km}\,\hbox{s}^{-1}\,\hbox{Mpc}^{-1}}

\newcommand{\ie}{{\it i.e.\ }}
\newcommand{\eg}{{\it e.g.\ }}

\newcommand{\hz}{Harrison-Zel'dovich }

\newcommand{\qrms}{\mbox{$Q_{rms-ps}$} }
\newcommand{\rms}{{\em rms} }

\newcommand{\be}{\begin{equation}}
\newcommand{\ee}{\end{equation}}
\newcommand{\cls}{\mbox{$C_l^{(S)}$}}
\newcommand{\clt}{\mbox{$C_l^{(T)}$}}

\def\deg{\ifmmode^\circ _\cdot\else$^\circ _ \cdot$\fi }    
\def\degg{\ifmmode^\circ \else$^\circ $\fi } 
\def \dtotrms{$\Delta T_{rms}/T$ }
\def\arcs{\ifmmode {'' }\else $'' $\fi}     
\def\arcm{\ifmmode {' }\else $' $\fi}     
\def\buildrel#1\over#2{\mathrel{\mathop{\null#2}\limits^{#1}}}
\def\mper{\ifmmode \buildrel m\over . \else $\buildrel m\over .$\fi}
\def\hper{\ifmmode \rlap.^{h}\else $\rlap{.}^h$\fi}
\def\sper{\ifmmode \rlap.^{s}\else $\rlap{.}^s$\fi}
\def\arcsper{\ifmmode \rlap.{' }\else $\rlap{.}' $\fi}
\def\arcmper{\ifmmode \rlap.{'' }\else $\rlap{.}'' $\fi}

\def\et{{\it et~al.~}}
\newcommand{\lta}{{\small\raisebox{-0.6ex}{$\,\stackrel
{\raisebox{-.2ex}{$\textstyle <$}}{\sim}\,$}}}
\newcommand{\gta}{{\small\raisebox{-0.6ex}{$\,\stackrel
{\raisebox{-.2ex}{$\textstyle >$}}{\sim}\,$}}}

\def\apj{ApJ}  
\def\mn{MNRAS}      

\title[Studies of CMB structure at Dec=+40\degg]{Studies of CMB
structure at Dec=+40\degg. \\ II: Analysis and cosmological
interpretation}
\author[Hancock et al.]
       {S. Hancock$^{1}$, C.M. Gutierrez$^{2}$, R.D. Davies$^{3}$,
A.N. Lasenby$^{1}$, G. Rocha$^{1}$, \and
R. Rebolo$^{2}$, R.A. Watson$^{2,3}$ and M. Tegmark$^{4}$\\
  $^{1}$Mullard Radio Astronomy Observatory, Cavendish Laboratory, 
        Madingley Road, Cambridge CB3 OHE, UK\\
 $^{2}$Instituto de Astrof\'\i sica de Canarias, 38200 La Laguna,
        Tenerife, Spain \\
  $^{3}$University of Manchester, Nuffield Radio Astronomy 
        Laboratories, Jodrell Bank, Macclesfield, Cheshire, SK11 9DL, UK\\
  $^{4}$Max-Planck-Institut f\"{u}r Physik, F\"{o}hringer Ring, D-80805
M\"{u}nchen, Germany
  }
\date{}
\begin{document}
\maketitle
\begin{abstract}

We present a detailed analysis of the cosmic microwave background
structure in the Tenerife Dec=+40\degg~data. The effect of local
atmospheric contributions on the derived fluctuation amplitude is
considered, resulting in an improved separation of the intrinsic CMB
signal from noise. Our analysis demonstrates the existence of common
structure in independent data scans at 15 and 33 GHz. For the case of
fluctuations des\-cribed by a Gaussian auto-correlation function, a
likelihood analysis of our combined results at 15 and 33 GHz implies an intrinsic \rms
fluctuation level of $48^{+21}_{-15}$ $\mu$K on a coherence scale of
$4\dg$; the equivalent analysis for a \hz model gives a power spectrum
normalisation of $\qrms = 22^{+10}_{-6}$ $\mu$K. The fluctuation
amplitude is seen to be consistent at the 68 \%  confidence level with that
reported for the COBE two-year data for primordial fluctuations
described by a power law model with a spectral index in the range $1.0
\le n \le 1.6$. This limit favours the large scale CMB anisotropy
being dominated by scalar fluctuations rather than tensor modes from a
gravitational wave background.
The large scale
Tenerife and COBE results are considered in conjunction with
observational results from medium scale experiments in order to place
improved
limits on the fluctuation spectral index; we find $n=1.10 \pm 0.10$
assuming standard CDM with $\ho=50 \kmspmpc$.
\end{abstract}


\section{Introduction}

Observations of fluctuations in the Cosmic Microwave Background (CMB) have been widely
recognized to be of fundamental significance to cosmology,
offering a unique insight into the physical conditions in the early
Universe. The amplitudes and distribution of such fluctuations provide
critical tests of the origin of the initial perturbations from which
the structures seen today have formed. On scales \simgt few degrees,
CMB observations probe scales of 1000's of Mpc, inaccessible to
conventional astronomy. At these large angles, the structures form part
of an intrinsic spectrum of fluctuations generated through topological
defects or inflation. In this linear growth regime, observations of the
scalar CMB fluctuations provide a clean measure of the normalisation of
the intrinsic fluctuation power spectrum.  This normalisation has been
established by a number of independent CMB observations 
(Smoot \et 1992, Ganga \et 1993, Hancock \et 1994).
In many theories, tensor CMB fluctuations from a background of
gravitational waves are also expected to be significant on these large
scales, and measuring the slope of the power spectrum offers the
potential to constrain this contribution to the CMB anisotropy (Hancock
\et 1994, Steinhardt 1993, Crittenden \et 1993). A comparison of the
large-scale anisotropy results with those on medium scales can also
be used to separate the scalar and tensor components under the assumption
of a specific cosmological model.

The Tenerife CMB experiments were initiated in 1984, with the
installation of the first 10 GHz switched-beam radiometer system at the
Teide Observatory on Tenerife Island. A subsequent programme of
development has led to the present trio of independent instruments
working at 10, 15 and 33 GHz. The ultimate objective is to obtain three
fully sampled sky maps covering some $\sim 5000$ square degrees of the
sky at each frequency and attaining a sensitivity of $\sim 50$ $\mu$K
at 10 GHz, and $\sim 20$ $\mu$K in the two highest frequency channels.
Drift scan observations have been conducted over a number of years
covering the sky area between Dec=+30\degg~and +45\degg.  The deepest
integrations have been conducted in the Dec=+40\degg~region and
resulted in strong evidence for the presence of individual CMB features
(Hancock \et 1994).

Davies \et 1995 (hereafter Paper I) described the performance of the
experiments and gave an assessment of the atmospheric and foreground
contributions to our data at Dec=+40\degg; here we analyse in detail
the results and cosmological implications of such observations. Section
2 describes the observational strategy and presents the stacked scans
at each frequency. In Section 3 we use several statistical methods to
calculate the level of the detected signals and their origin. A
statistical comparison with the results of the COBE DMR two-year data
is conducted in Section 4 and used to place limits on the spectral index
of the primordial fluctuations. In Section 5 we use the additional information
provided by medium-scale anisotropy results to provide improved limits
on $n$.

\section{The scans at Dec $+40\dg$}

\subsection{Observations}

Observations were conducted at the three frequencies 10, 15 and 33 GHz
by drift scanning in right ascension at a fixed declination of
40\degg. The measurements were made independently at each frequency,
using separate dual-beam radiometer systems as described in Paper I.
The three instruments are physically scaled so as to produce
approximately the same beam pattern (FWHM$\sim 5\degg$) on the sky,
thus allowing a direct comparison of structure between frequencies. A
characteristic triple beam profile (switching angle 8\deg 1) is
obtained by the combination of fast switching (63 Hz) of the horns
between two independent receivers and secondary switching
(0.125 Hz) provided by a wagging mirror. We make repeated observations
of the sky, binning the data in 1\degg~intervals in RA and stacking them
together in order to reduce the noise as compared with individual
measurements. As a consequence of using two independent channels, the
receiver noise contribution to the final data scans is reduced by a
factor $\sqrt{2}$ compared with single channel observations.

The data considered in this paper are the same as those presented in
the preliminary report by Hancock \ea (1994). As before we restrict our
analysis to the RA range $161\dg-230\dg$ corresponding to Galactic
latitude $b>56\dg$. At these high latitudes foreground emission from
the Galaxy is expected to be at a minimum. This sky region has also been
selected (see Paper I) to be free from discrete radio sources above the 1.5 Jy
level at 10 GHz; assuming a flat spectrum this corresponds to an
antenna temperature contribution of 48, 21 and 5 $\mu$K at 10, 15 and
33 GHz respectively in the Dec=+40\degg~data.
The actual contamination is reduced below these levels by using the K\"{u}hr catalogue (K\"{u}hr \ea 1981)
supplemented by the VLA calibrators list to subtract the
discrete source contribution from each data scan.
This amounts to making corrections to the
\rms signal level of 13, 8, and 3 $\mu$K at 10, 15 and 33 GHz respectively
The contribution of
unresolved radio sources is expected to be $\Delta T/T \le 5\times
10^{-6}$ at 15 GHz and significantly smaller at 33 GHz (Aizu \et 1987,
Franceschini \et 1989).

Figure 1 presents our final stacked scans at 10, 15, 33 GHz and the
weighted addition (named 15+33) of the data at the two higher
frequencies. We have calculated the standard error-bars of each point
considering the presence of correlated atmospheric
noise between the two channels
of each receiver system as described in Paper I. This represents a mean
enhancement of 3 \% at 10 GHz, 4 \% at 15 GHz, and 16 \% at 33 GHz with
respect to the estimation of the noise computed in Hancock \et (1994).
The
revised sensitivities per beam-sized area are 61, 32, 25 and 20 $\mu$K
at 10, 15, 33 and 15+33 respectively. A visual inspection reveals the
presence of common features in the scans at the two higher frequencies;
this is supported by a cross-correlation analysis in Section 3.2.

\begin{figure}
\hspace*{-6in}
\includegraphics{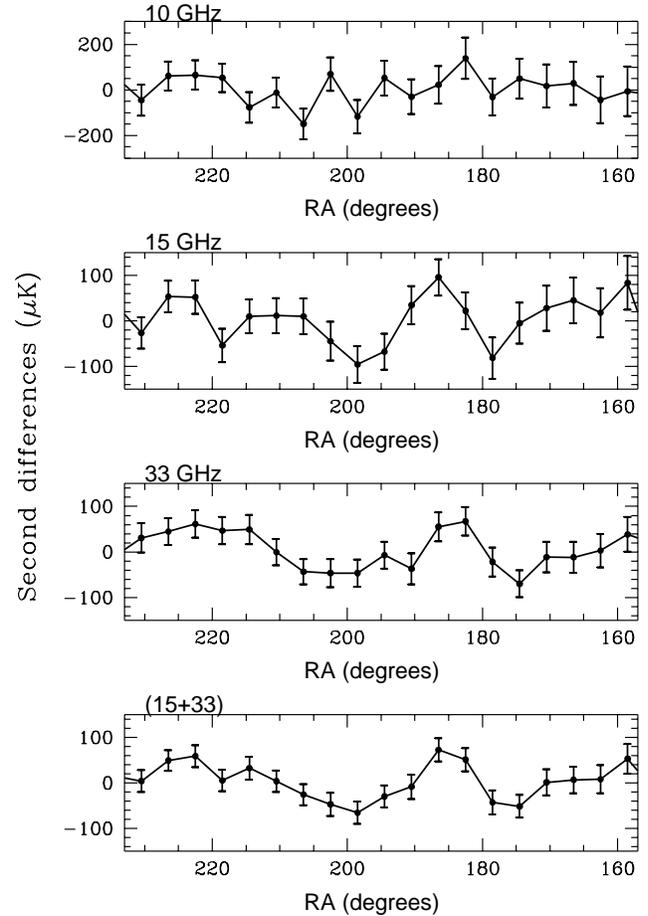}
\vspace*{13cm}
\caption{The stacked scans at Dec=+40\degg. The second
differences in temperature are shown binned at
$4\dg$ intervals in RA.
The error bars are at 68 \% confidence and take into
account any atmospheric correlations between receiver channels. 
The scans labeled 15+33 represent
the weighted addition of the data at 15 and 33 GHz. }
\label{Figure1}
\end{figure}

\subsection{Reliability of the detected signals}

Determination of the amplitude of the CMB component of the structure
requires one to consider the contributions of random noise and
foreground signals to the observed data scans. The former has its
origin in the thermal variations in the receivers and in the
fluctuating component of the atmosphere, whilst the latter consists
primarily of free-free and synchrotron emission in the Galaxy, plus
emission from the Sun and Moon. Of these effects, only the Galactic
emission remains constant from day to day at a given frequency. In
Paper I we have estimated a maximum Galactic contribution of $\Delta T_{rms}
=4 \mu$K
in the results at 33 GHz. An improved separation between the Galactic
and the cosmological signal at each frequency will be presented in a
forthcoming paper. We have removed the effects of the Sun and Moon to
better than 50 dB in each daily scan, and stacking $n$ days of
observations with the Sun and Moon in different positions will reduce
any residual contribution to an insignificant level.

The thermal noise contribution from the receivers conforms to a random
Gaussian distribution and integrates down accordingly. In each
instrument, the channel 1 receiver is fully independent of the channel
2 receiver and hence there is no correlation between the receiver noise
in the data recorded for the two channels.  Atmospheric emission varies
randomly over days and thus at each frequency, stacking together $n$
days of observations reduces the atmospheric signals in a given day by
$\sqrt{n}$. Additionally, the Tenerife observing strategy is such that
the {\em independent} 10, 15 and 33 GHz instruments observe {\em
different} declination strips on any given day so that the atmospheric
contribution to data taken at each frequency is uncorrelated. This
offers a key advantage over experiments such as Saskatoon (Wollack \et
1993), where the data in all of the frequency channels are highly
correlated thus introducing an element of uncertainty due to the need
to account for this. In our individual measurements the atmospheric
contribution to each channel is correlated, since the timescale of the
receiver switching is short compared with that for typical atmospheric
fluctuations.  In previous papers (Davies \ea 1987, Watson \ea 1992,
Hancock \ea 1994 and Guti\'errez \ea 1995) this effect was not
considered. The presence of a correlated atmospheric component affects the
estimation of
the astronomical signal based on the difference of the $(A+B)/2$ and
$(A-B)/2$ data sets ($A$ and $B$ being separate subsets of the data).
In particular if $A$ and $B$ each correspond to one
channel, $(A+B)/2$ contains the contribution of the astronomical
signal, and the atmospheric and instrumental noise, whilst $(A-B)/2$
only contains the instrumental noise. This is the case for
the analysis of the 33 GHz data presented in
Hancock \et (1994) and therefore the signal level obtained from the
difference in variance between the $(A+B)/2$ and $(A-B)/2$ scans
contains a contribution from the atmosphere as well as the astronomy.
Note that at 15 GHz the data were subdivided according to the
observing epoch and consequently the contribution of the atmosphere to
the derived signal was largely reduced.

Here we present a new split of the 33 GHz data into two subsets
$X$ and $Y$ such that both channels of a given scan are included
in the same data subset. Considering the non-repeatiblity of the atmosphere
from day to day common atmospheric signals are not expected to occur
in both subsets. Also, considering the large number of independent
observations (more than 50 for each channel)  we expect that the net
effect of the atmospheric signals will be just an increase of the
variance in the final stacked scans of each subset, this increase in
variance being
approximately the same in both. For these reasons, when $X$ and $Y$ are
combined to form the $(X+Y)/2$ and the $(X-Y)/2$ scans we expect to
have approximately the same atmospheric signals in the sum and in
the difference. Figure 2 shows the stacked data scans for the $X$ and $Y$
subsets and their sum and
difference. We see a general agreement between the results of splits
$X$ and $Y$, and the presence of common features in both. An analysis
similar to that presented in Hancock \ea (1994) gives an astronomical signal
($\sigma _s^2=\sigma_{(X+Y)/2}^2-\sigma_{(X-Y)/2}^2$) with an amplitude
$\sigma _s=43 \pm 12$ $\mu$K.
The value of the signal quoted in Hancock \et
was $\sigma_{old}=49 \pm 10$ $\mu$K; the difference with our improved
estimation is certainly due to the subtraction of the atmospheric
signal. The difference between both estimates ($\sigma
=(\sigma_{old}^2-\sigma _s^2)^{1/2}$) is $\sim 23$ $\mu$K
which is our best
assessment of the atmospheric contamination in the analysis based on
the addition and difference of the 33 GHz data; this value is in good
agreement with estimates obtained using other methods.

\begin{figure}
\psfig{file=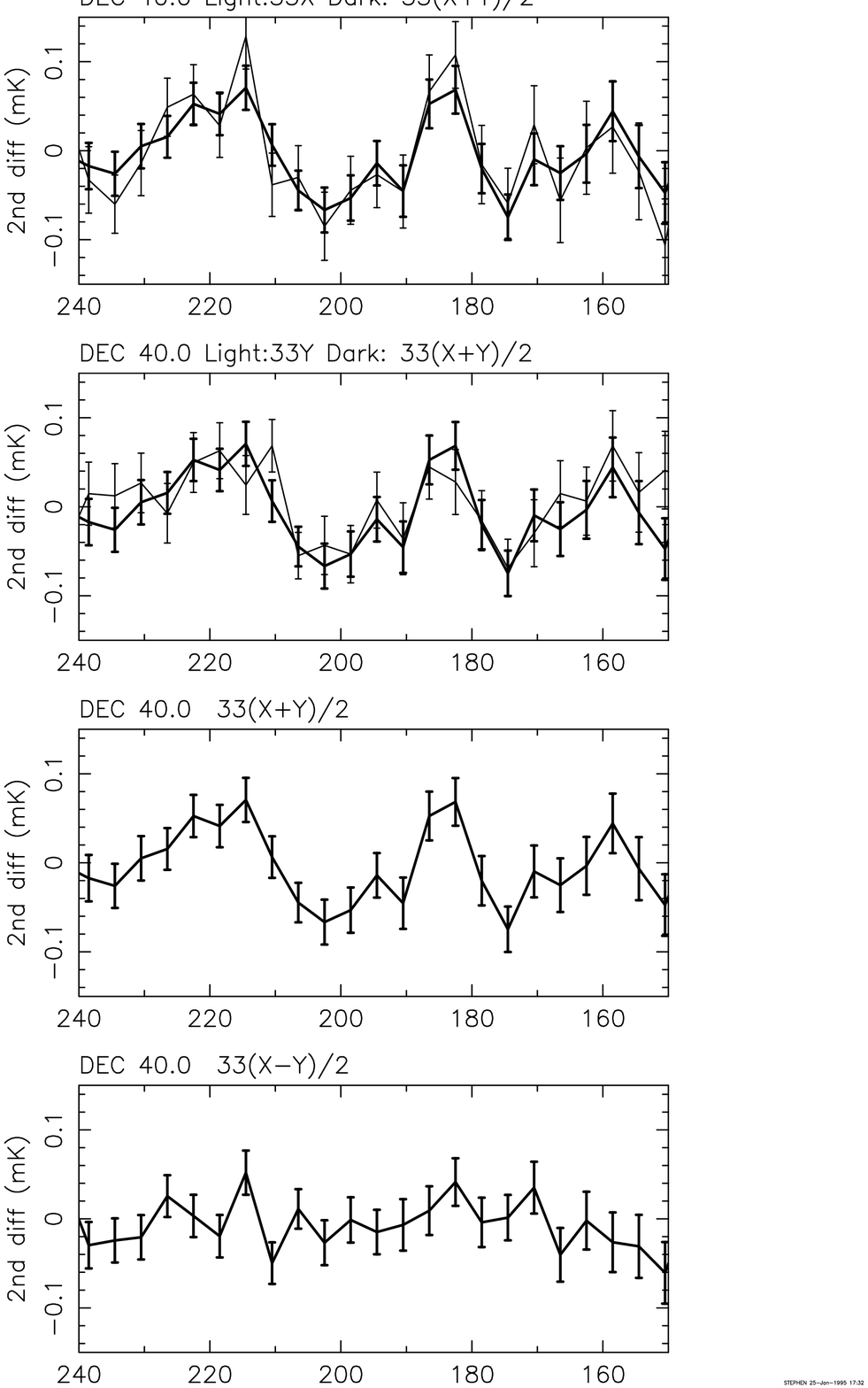,width=3in,clip=}
\caption{The data scans obtained from the analysis of two subsets
of the 33 GHz data. In a) and b) are shown
the subsets denoted X and Y which are constructed
such that the atmospheric contributions to each are independent:
see main text for details. The differences in the 
$(X+Y)/2$ (panel c)) and $(X-Y)/2$ (panel d)) scans should therefore be
due only to astronomical signals.}
\label{Fugure2}
\end{figure}

\section{Statistical analysis} 
The estimates of the astronomical signals made
in the previous section can be improved on using a detailed statistical
analysis. Here we use a likelihood analysis and
take into account the contribution of the correlated atmospheric noise
by enlarging the error bars on the scans as shown in Figure 1.
These modified scans are also used to study the presence of
common features at 15 and 33 GHz by the calculation of the
cross-correlation function.

\subsection{Likelihood analysis}
\label{likelihood}

This analysis takes into account all the relevant parameters of the
observations: experimental configuration, sampling, binning, etc.  The
combination of the atmospheric and instrumental noise can be modeled by
a Gaussian distribution uncorrelated from point to point, (see previous
section), which implies that the noise only contributes to the diagonal
terms of the covariance matrix. We also assume that the astronomical
signal is described by a Gaussian random field and therefore our
results correspond to a superposition of Gaussian fields in which all
their statistical properties are specified by the covariance matrix,
which takes into account the full correlation between the data points.

We have analyzed our results for two hypothetical sky models, the first
of which corresponds to a signal described by a Gaussian
auto-correlation function (ACF) with amplitude $\sqrt{C_0}$ and width
$\theta _c$. This is not a realistic physical scenario but has been
used widely in the past (Davies \et 1987, Readhead \et 1989, Watson \et
1992) because it provides for an easy comparison between the results of
experiments with different configurations. The intrinsic ACF for these
models is given by

\begin{equation}
C(\theta)=C_0\exp (-\frac{\theta ^2}{2\theta _c ^2}).
\end{equation}
which is modified accordingly by our triple beam filtering (Watson \et
1992).  Our instrument is sensitive over a range of coherence angles $
1\dg \simlt \theta_c \simlt 10\dg$, attaining peak sensitivity for a
coherence angle of 4\degg. We have analyzed the $X$ and $Y$ subsets for
15 and 33 GHz, the total stacked scans at these two frequencies, and
our best scan 15+33. The results for $\theta_c=4\dg$ are presented in
the third column of Table~1.  The amplitude of the intrinsic signal
corresponding to the maximum likelihood is given, along with the
one-sigma confidence bounds calculated in a Bayesian sense with uniform
prior. All results look consistent with clear detections at the two to
three sigma level, and mean values of the signal slightly smaller than
those presented in Hancock \et, due to our improved estimate
of the error bars in the stacked scans (see Section 2 and Paper I).
Figure 3 presents the contours of equal probability for the 15+33 scan.
We see a well defined point of maximum likelihood at $\theta
_c\sim 4$\degg~and $\sqrt{C_0}\sim 50$ $\mu$K.

The second model considered here is more interesting from a
cosmological viewpoint. It corresponds to the prediction of the power
law form ($P(k)\propto k^n$) for the spectrum of the primordial
fluctuations. Considering only the Sachs-Wolfe part of the spectrum of
the fluctuations the intrinsic ACF can be expressed as
\begin{eqnarray}
C(\theta)=\frac{\qrms^2}{5}\sum _l (2l+1) C_l^{(S)} P_l(\cos \theta)
\label{eq:cl}
\\
C_l^{(S)}=C_2^{(S)}
\frac{\Gamma[l+(n-1)/2]\;\Gamma[(9-n)/2]}{\Gamma[l+(5-n)/2]\;\Gamma
[(3+n)/2]} \nonumber
\end{eqnarray}
where the sum is extended to the multipoles $l\lta 60$ which
corresponds to the range of angular sensitivity of our experiments. 
For $l \gta 20$, standard models predict additional contributions to the CMB 
anisotropy, as one moves into the low $l$ tail of the CMB Doppler peak.
Hence fitting for the Sachs-Wolfe term alone (Equation~2) to CMB data on these
scales can lead to the derived values for $n$ being increased by as much as
10\% over the true primordial value. This point should be borne in mind when
comparing the limits on $n$ from the Sachs-Wolfe term (Sections~3 and~4) to
those from a fit to a full CDM type functional form as in Section 5.

For
a given value of the spectral index $n$, the intrinsic ACF is a
function only of \qrms.
Figure 4 shows the likelihood surface as a function of \qrms and $n$
for the 15+33 scan.
The peak likelihood forms a ridge displaced from zero
in \qrms and corresponds to a $3-4$ sigma detection of structure
for each value of $n$ considered. The shape of the surface implies
that all values of $n$ in this range are equally likely. 
This is predominantly
a consequence of our observing technique which samples
only a small angular range of the spectrum of fluctuations. Thus whilst
our observations provide a good measure of the fluctuation power on
$\sim 4\dg$ scales, they do not in themselves contain sufficient
information about the distribution of power with angular scale to allow a useful
determination of the spectral slope: for this one must compare with
experiments on other angular scales (see Sections 4 and 5).  For the
specific case of a \hz spectrum ($n=1$)
the results of the likelihood
analysis are given in column two of Table~1. The normalisation \qrms
corresponds to the maximum of the likelihood function and the
confidence intervals are at 68 \%, calculated in the standard Bayesian
manner using uniform prior. We see that in general the results are
consistent and agree with a global normalization of the quadrupole
$\qrms \sim 20-25$ $\mu$K. Our best estimate for \qrms of $22^{+10}_{-6}
\mu$K from the 15+33 scan is reduced over the value of $26\pm 6 \mu$K
previously reported due to our now having properly accounted for the correlated
atmospheric noise.
\begin{table}
\begin{center}
\caption{Results of the likelihood analysis for a Harrison-Zel'dovich
spectrum of fluctuations (second column) and for a Gaussian ACF (third
column).}
\vskip 0.2 cm
\begin{tabular}{ccc}
& &  \\
$\nu$ (GHz) &  \qrms ($\mu$K) & $\sqrt{C_0}$ ($\mu$K) \\
\hline 
& & \\

15A & $27^{+16}_{-16}$ & $54^{+37}_{-30}$ \\
& & \\
 
15B & $17^{+9}_{-17}$ & $24^{+21}_{-24}$ \\
& & \\

15  & $21^{+12}_{-9}$  & $44^{+26}_{-19}$\\
& & \\

33A & $22^{+14}_{-10}$ & $45^{+32}_{-24}$ \\
& & \\

33B & $28^{+12}_{-9}$ & $57^{+28}_{-25}$ \\
& & \\

33  & $24^{+11}_{-8}$ & $49^{+27}_{-17}$ \\
& & \\

15+33 & $22^{+10}_{-6}$ & $48^{+21}_{-15}$ \\
& & \\

\end{tabular}
\end{center}
\end{table}

\begin{figure}
\psfig{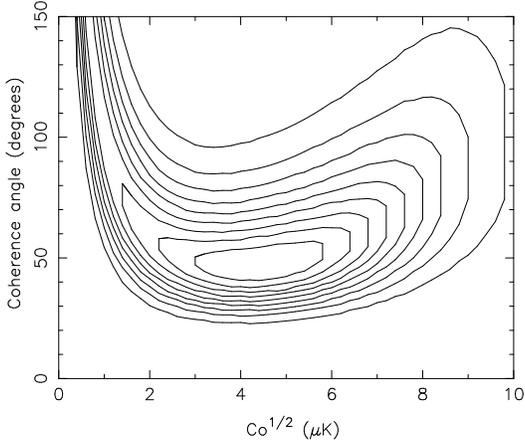}
\caption{Contour levels of equal likelihood for the 15+33 scan in the
case of a Gaussian shaped ACF. The contours correspond to 10, 20, 30 ...
90 \% of the total probability distribution.
The Tenerife configuration obtains maximum sensitivity for coherence
angles in the range $2\dg \simlt \theta \simlt 6\dg$;
structure is clearly detected at $\sim 50 \mu$K over this angular scale
range.}
\label{Figure3}
\end{figure}

\begin{figure}
\hspace*{-0.36in}
\psfig{file=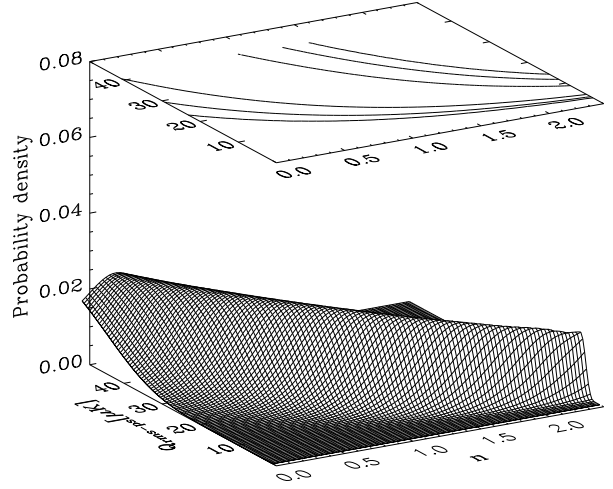,width=4in} 
\caption{The two-dimensional normalised likelihood surface as a
function of the spectral index $n$ and the normalisation \qrms
for the Tenerife data. The projected contours are at 68 \%, 95 \% and
99 \% confidence.}
\label{Figure4}
\end{figure}

\subsection{Cross-correlation analysis}
\label{ccf}

We have investigated the presence of common features in our $15+33$
scans at
Dec=+40\degg~by evaluating the cross-correlation function,
\begin{equation} C(\theta)=\frac{\sum _{i,j} \Delta T_i\Delta
T_j^{\prime} w_iw_j^{\prime}}{\sum_{i,j} w_iw_j^{\prime}} \end
{equation} where $\Delta T$ and $\Delta T^{\prime}$ denote the second
differences in the stacked scans at 15 and 33 GHz respectively. The sum
is extended over all points subtending a mutual angle $\theta$ in RA,
with weights $w_i=1/\sigma _i^2$ and $w_j^{\prime}=1/\sigma^{\prime
2}_j$ respectively.  The data points in Figure 5 represent the mean and
68 \% confidence intervals of $C(\theta)$. The observed profile is
characteristic of the triple beam sampling and it is clear that for
angles $\theta \lta 20\degg$ the results are consistent with the
presence of common signals in the two independent data sets.  Because
of the non-independence of the data points in the cross-correlation
function, the statistical significance is difficult to evaluate,
requiring the computation of the covariance matrix of the correlation
function (see \eg Ganga \et
1993).  However, considering only the correlation at zero lag, the
observed value of $C(\theta)=1100^{+680}_{-720}$ $\mu$K$^2$ is found to be
inconsistent with noise at the 95 \% level. This value for the
amplitude of the component of the signal common to the 15 and 33 GHz
data is in agreement with the equivalent estimates of the variance of
the signals present in each data scan separately, demonstrating that
common signals can account for all of
the structure seen in the separate scans. This is contrary to what one
expects for structure of Galactic origin, which in the case of
synchrotron and free-free emission would exhibit a change in amplitude
by a factor 9 and 5 respectively on moving from 15 GHz to the higher
frequency of 33 GHz. This result taken in conjuction with
the limits on the Galactic contamination of the 33 GHz scan (see
above) and the successful comparison of the features seen in the 15 and 33
GHz scans with the higher frequency COBE data (Lineweaver \ea, 1995),
leads us to assign a cosmological origin to the majority of the
signal present in the 15 GHz data.

We have explicitly tested the consistency of the common signals with
fluctuations originating in inflationary type models which predict
an approximately scale invariant $n\simeq 1$ spectrum of fluctuations.
Using the normalisation of $\qrms = 22 \mu$K as
derived from the likelihood analysis of the 15+33 scan we simulated
5000 realisations of a sky conforming to a \hz spectrum and repeated
the Tenerife sampling strategy. The mean value (solid line) and
one-sigma confidence intervals (dashed lines) of the auto-correlation
function obtained from these realisations are plotted in Figure 5.
There is clearly a good level of agreement in the amplitude and
shape of the experimental results and the theoretical model.
Statistically the Tenerife observations have the properties expected for
primordial fluctuations generated from inflation, and although these
results do not prove inflationary theory, this does offer a plausible
explanation for the existence of structure on scales greater than the horizon
volume at recombination. In this case, the clearly defined structures
visible in the scans in Figure 1 represent the seed perturbations
generated in the inflationary era at $t<10^{-34}$s. In the following section
we investigate the potential to prove or disprove inflation by measuring
the slope $n$ of this primordial spectrum of fluctuations using large
scale CMB anisotropy measurements.

\section{Statistical comparison with COBE DMR}
\label{comb}
A comparison between the results of different CMB experiments offers
the opportunity to check independent measurements, to extend the range
in frequency and angular scale, to constrain cosmological models and,
if the sensitivity of the experiments is sufficient, to compare
features. There are several experiments operating on angular scales of
a few degrees: MIT (Ganga \et 1993), COBE (Smoot \et 1992, Bennett \et
1994), RELIKT (Strukov \et 1993), ARGO (De Bernardis \et 1994a)
and Tenerife. The
first comparison between independent CMB observations
was made by Ganga \et (1993) who  found a
clear correlation between the results of the first year of COBE DMR
observations and those of the MIT experiment. De Bernardis \et (1994b)
have also made a statistical comparison between the amplitude of the
signal reported for ARGO and that of the COBE DMR first year results,
from which they constrain the spectral index to be
$0.5\le n \le 1.2$ in the absence of any gravity wave background.
In our preliminary report (Hancock \et 1994)
we compared the amplitude of the signal detected on $\sim 5\dg$ scales
in our Dec=+40\degg~data with that on $\sim 7\dg$ scales for the first
year of COBE data. We found that both results were consistent with
inflationary models ($n\simgt 0.9$)
but with a favoured spectral index of $n=1.6$.
Lineweaver \et (1995) have presented the first direct comparison of CMB
features between the two-year COBE DMR data and the Tenerife
Dec=+40\degg~observations, confirming the agreement in the level of the
normalization of both experiments and providing clear evidence for
the presence of common hot and
cold spots in both data sets. The comparison presented here is
different to that  conducted in Hancock \et (1994) in that we use
a more rigorous comparison technique that utilises the likelihood
function to incorporate fully the effects of cosmic and sample variance,
random noise and the interdependence of the model parameters.
In addition to the improvements from this revised analysis, our new
results also reflect the increased sensitivity of the COBE data after
two years of observing, along with the more accurate estimate of
the cosmological signal in the Tenerife data.

\subsection{Properties of the two data sets}
The instrumental profile of the COBE data is described approximately by a
Gaussian beam with FWHM$\sim 7\degg$. As can be seen in Figure 1 of Watson
\et (1992)
there is a range of angular scales to which the COBE DMR and
Tenerife experiments are both sensitive. Measurements taken by COBE
cover the full sky, but to determine the CMB fluctuations the region of
the Galactic plane has been excluded  ($|b|\le 20$\degg) thereby
introducing a degree of uncertainty in estimating the properties of the
global field; this effect is commonly termed sample variance
(Scott, Srednicki \& White 1994). The
uncertainties in the COBE two year
results are dominated by the effect of cosmic
variance {\it i.e.} the fact that our stochastic theory describes the
Universe as a particular realisation of a random field.
Together the cosmic and sample uncertainties form an intrinsic limitation
of the COBE experiment since unlike random errors they
are not reduced by increased integration time.
The two year COBE data have been analysed independently by a number
of authors (\eg Banday \ea 1994, Bennett \ea 1994, Bond 1994, Gorski \ea 1994,
Wright \ea 1994). All find evidence for statistically significant structure
at an amplitude consistent with that of $30\pm5$ $\mu$K \rms on a $10\dg$
scale announced by Smoot \ea (1992) for the first year data.
The best fit values for $n$ and \qrms depend on the precise analysis
techniques employed, but are generally consistent with the
values of $n=1.10 \pm 0.29$,
$\qrms=20.3\pm 4.6$ $\mu$K found by Tegmark and Bunn (1995) for the combined
53 and 90 GHz data with
the quadrupole included.
In the case of the Tenerife experiment the
double-switching scheme removes the contribution of low order
multipoles decreasing the cosmic variance of the signal on these
large scales; the major
source of uncertainty is produced by the partial sky coverage (sample
variance) and the instrumental noise. The region observed by the
Tenerife experiments covers $\sim 5000$ square degrees but here we have
limited our analysis to our region of deepest integration at high
Galactic latitude which constitutes a sample $\sim 500$ square degrees.
For such a region the uncertainties due to the partial sky coverage
dominate over the intrinsic variance by a factor $\sim 10$ (Scott,
Srednicki \& White
1994) and the combined uncertainty is approximately of the order of
that introduced by the instrumental noise in the 15+33 scan.

In previous work (Hancock \ea 1994) we explicitly took into account the
effects of cosmic and sample variance by using Monte Carlo simulations.
This was necessary because the simple excess variance statistic used in
the comparison incorporated only the uncertainty due to random noise.
The comparison by necessity assigned equal probability
to all values of $n$ for the COBE data, since the COBE first year
results were only published in the form of a best fit $n$ versus \qrms
relation rather than a full two dimensional probability distribution.
What is required is a data analysis technique that allows the joint
probability of any combination of the model parameters to be calculated
and which implicitly takes into account random errors and cosmic and
sample uncertainties.  The Bayesian approach using the likelihood
function as described in Section 3 attempts to do precisely this.  The
likelihood function peaks at the most likely parameters (the best estimate
of the true
values if the likelihood function is unbiased) and has some
distribution which through Bayes theorem is representative of the
combined effects of the cosmic, sample and random uncertainties.  The
issue of how well this distribution reflects the true uncertainties is
addressed explicitly in a forthcoming paper (Rocha \ea in preparation)
by comparison of the Bayesian probability distribution with that
obtained from direct Monte Carlo simulations of the data. The Bayesian
and frequentist approaches are found to be consistent for the Tenerife
data and to a good approximation
the likelihood function is also seen to be an unbiased
estimator of the model parameters.  Consequently the likelihood surface
for the joint Tenerife and COBE data set provides the definitive means
of comparison of the observations under some assumed sky model.

\subsection{The Tenerife-COBE likelihood function}
\label{likejoint}
Here we apply the likelihood analysis to the COBE two year data and the
Tenerife 15+33 scan, assuming a power law model with free parameters
$n$ and $\qrms$. The COBE Galaxy-cut two-year map consists of 4038
pixels, whilst the Tenerife Galaxy cut (RA $161\dg-230\dg$) scan contains 70
pixels, requiring a 4108 $\times$ 4108 covariance matrix for a joint
likelihood analysis of the data. The direct inversion of such a large
matrix, necessary for the likelihood analysis, is computationally
intensive, but has been implemented for the combined 53 and 90 GHz COBE
two-year data by Tegmark and Bunn (1995), hereafter ``TB95". 
A number of other authors
(Bond 1994, Bunn and Sugiyama 1994, Gorski 1994) have computed the
likelihood function by using various 
data compression techniques to reduce the
size of the covariance matrix. The compression is achieved by
discarding noisy data vectors whilst retaining most of the cosmological
signal, and although the results are close to optimal, the slight loss of
signal results in marginally larger error bars than the ``brute
force'' approach.
The latter method is conceptually the simplest one, since it
is merely a likelihood analysis using all available data, and involves
no adjustable parameters such as the degree of data compression.
Since each matrix inversion requires merely about 10 minutes on a
fast workstation, we use the brute force method here.

We arrange the pixels in a 4108-dimensional vector and compute the 
likelihood function as in TB95 by Cholesky
decomposition of the 4108 $\times$ 4108 covariance matrix at a dense grid
of points in the $(n,\qrms)$-parameter space, marginalizing over
the four ``nuisance parameters" that describe the monopole and dipole.
The covariance matrix consists of three parts: a $70 \times 70$ block with the
covariance between the Tenerife pixels, a $4038 \times 4038$ 
block with the covariance between the COBE pixels,
and off-diagonal $4038 \times 70$ blocks containing the covariance between 
the Tenerife and COBE pixels. 
Simply multiplying the likelihood curves resulting from 
two separate analyses of the COBE and Tenerife data sets would correspond 
to neglecting the off-diagonal blocks, and this is clearly only a good
approximation if the two data sets are almost uncorrelated. We find that
the inclusion of the cross-terms makes a non-negligible difference, which 
is not surprising in view of the fact that the two experiments 
probe comparable angular scales
and have observed a common region of the sky.

For our Tenerife data, the best estimate of the cosmological signal is
obtained from the 15+33 combined scan after correction of the error
bars for the correlated atmospheric noise term.  The possible
contribution of Galactic signals has been estimated from the 10 GHz
data to be less than 4 $\mu$K at 33 GHz and has not been considered in
the current comparison. The normalised likelihood function for this
scan, as plotted in Figure 6, represents the joint probability of
obtaining a given combination of $n$ and $\qrms$. On its own, the Tenerife
configuration provides less leverage on the slope of the spectrum than
the COBE satellite. This is because the Tenerife experiment is 
insensitive to the largest angular scales, and because
the one-dimensional shape of the 
dec$+40^\circ$ strip makes it 
difficult to separate the power contributions from different 
scales. In other words, a narrow strip corresponds to
wide window functions in $\ell$-space, with considerable
aliasing of small-scale power onto larger scales
(just as the case is with one-dimensional ``pencil beam" 
galaxy surveys). 
As a result, the Tenerife data can be equally well
fit by a range of $n$ and $\qrms$ values, 
resulting in a likelihood ridge in $(n,\qrms)$-space
with minimal discriminatory power for
the parameter $n$. In contrast, the COBE observations are
sensitive to the slope to the extent that the likelihood surface is
peaked in the $n$-dimension. 
Combining the COBE information with the
Tenerife data improves the situation in two ways:
it extends the lever arm on the spectral slope from the COBE scales down to
the $4\dg$ scale of Tenerife, and in addition eliminates the above-mentioned
aliasing problem, since the joint data set is no longer 
one-dimensional.

\begin{figure}
\hspace*{-1.in}
\psfig{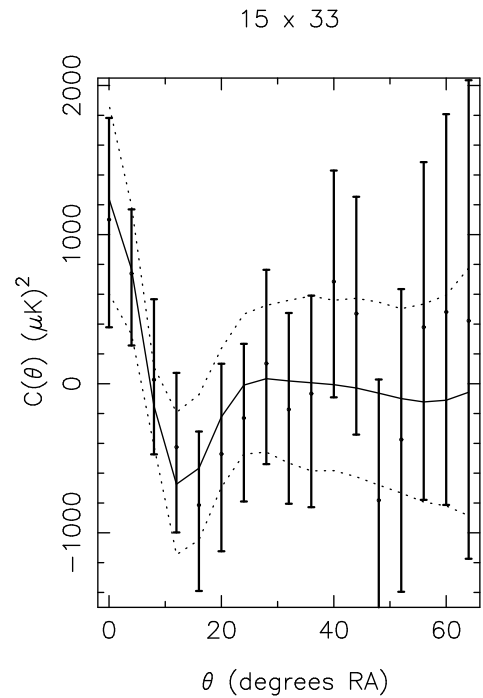}
\caption{The cross-correlation function $C(\theta)$ for the 15 GHz and
33 GHz data over RA=161\degg$-$230\degg~showing the data points and the
one sigma  error-bars. The solid line corresponds to the prediction of
a Harrison-Zel'dovich spectrum of fluctuations with a normalization of
$\qrms=22$ $\mu$K. The dashed lines are the 1$\sigma$ confidence bounds
arising from cosmic variance and sample variance.}
\label{Figure5}
\end{figure}

Figure 6 shows the confidence contours obtained from
Bayesian integration under the combined COBE and Tenerife likelihood
surface assuming a uniform prior. The 68\% joint confidence region in
$(n,\qrms)$-space encloses a region from 0.90 to 1.73 in $n$ for
$\qrms$ in the range from $12.1$ to $22.9$ $\mu$K, 
with the peak at $n=1.37$,
$\qrms=16.1\mu$K. Marginalizing over $\qrms$ with a uniform 
prior, one obtains the probability
distribution for $n$ as given in Figure 7,
corresponding to $n=1.33\pm 0.30$ at $68\%$ confidence.
The resulting limits on the normalization, conditioned on 
$n=1$ as is customary, are $\qrms=21.0\pm 1.6$. 
The corresponding results in TB95 using just the COBE data, and
including the weak correlated noise term (Lineweaver \ea 1995),
were $n=1.10\pm 0.29$ and $\qrms=20.3\pm 1.5$, with the peak likelihood
located at $n=1.15$, $\qrms=18.2 \mu$K.
In other words, although the total normalization has risen by 
a mere $3\%$, the slope estimate has risen by $12\%$ and the peak
likelihood has been shifted to higher $n$ and lower $\qrms$. 
This indicates that the higher angular 
resolution data from the Tenerife experiment contains slightly more
power on small scales. As explained in Section~3.1, this is not unexpected,
since the presence of a Doppler peak would cause a rise in the power spectrum
at higher $l$.


\begin{figure}
\hspace*{-0.35in}
\psfig{file=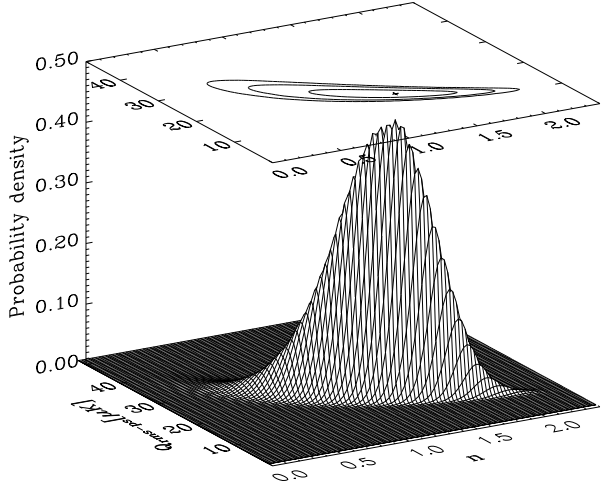,width=4in}
\caption{Constraints on the quadrupole \qrms and on the
spectral index $n$ of fluctuations obtained from a joint likelihood analysis of 
the Tenerife data and COBE DMR two-year data. The contour levels
represent 68 \%, 95 \% and 99 \% of the region of joint probability.
The peak of the distribution lies at $n=1.37$, $\qrms=16.1 \mu$K and is identified by the cross.}
\label{Figure6}
\end{figure}

\begin{figure}
\psfig{file=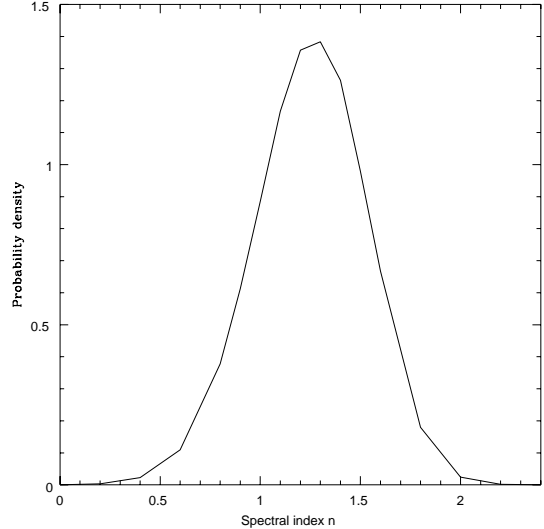,width=3in}
\caption{The marginal likelihood for the spectral index $n$ as
obtained from the joint analysis of the Tenerife and COBE data.
The spectral index is seen to lie in the range $1.02 \le n \le 1.62$ at
68\% confidence, with a best fit value of $n=1.33$.}
\label{Figure7}
\end{figure}

\section{Additional constraints on the primordial spectral index}

The large
angular scale CMB observations from Tenerife and COBE probe
fluctuations that have yet to go non-linear and the shape of the power
spectrum is thus insensitive to the exact abundance of the baryonic
mass ($\Omega_b$)
and to the value of $h=\ho/100\kmspmpc$. However, together
the Tenerife and COBE observations provide a direct measure of the
CMB power spectrum normalisation, against which one can compare
intermediate scale observations and hence discriminate between competing
cosmological models.
Here we stress the importance in the agreement between the derived normalisation
for the independent Tenerife and COBE experiments, which are
subject to different systematic errors and different
foreground contamination. The accuracy to which we know the
normalisation of the power spectrum
clearly becomes an issue when comparing with smaller
scale observations to determine cosmological parameters.
This is a particular concern if, as has been suggested (Crittenden
\ea 1993, Steinhardt 1993, Abbott and Wise 1984),
a component of the large scale anisotropy
may be due to tensor metric perturbations produced from a background of
gravitational waves. These can
arise naturally in inflationary scenarios and would contribute a component
$\clt=<|a_{lm}^{T}|^2>$ to the observed CMB angular power spectrum
(cf Equation \ref{eq:cl}).
The ratio of
the tensor modes \clt to the scalar modes \cls (primordial
density fluctuations) is highly suppressed for fluctuations
contained within the horizon volume at recombination and hence their
contribution is only significant on scales \simgt $2\dg$.
Consequently the existence of a tensor contribution has implications when
comparing the large scale anisotropy level with that on smaller scales
and with large scale structure
observations in order to test cosmological models. 
The anisotropy measurements from Tenerife and COBE fix the sum of
\clt and \cls, but the separation of the two terms requires a comparison with smaller scale observations under the  assumption
that a given cosmological model is correct (Crittenden \ea
1993, Steinhardt 1993). In the case of power law inflation 
a relation exists between the tensor to scalar ratio
$C_2^{(T)}/C_2^{(S)}$ and the slope of the primordial power spectrum
(Crittenden \ea 1993, Steinhardt 1993) :
\be
C_2^{(T)}/C_2^{(S)} \approx 7(1-n)
\label{eq:tensors}
\ee
from which we see that the two contributions are comparable at
$n=0.85$ with \clt decreasing relative to \cls for higher values of $n$.
Thus the limit of $n \ge 1.0$ obtained from the analysis of the
combined Tenerife and COBE data in Section \ref{likejoint} 
implies that it is unlikely that the tensor component will be dominant in such models.
In the remainder
of this section we shall investigate this in more detail by comparing with 
medium-scale anisotropy results.

Contemporary cosmological models with adiabatic fluctuations predict a sequence of peaks on the power spectrum which are generated by acoustic oscillations of the photon-baryon fluid at recombination. Of particular interest is the height and position of the main acoustic peak --- the so called Doppler peak: the height depends on quantities like the baryonic content of the universe, $\Omega_{b}$, and Hubble constant, $H_{0}$, whilst the position depends on the total density of the Universe, $\Omega_{0}$.   

The work required to place the existing medium scale anisotropy results 
into a common statistical framework and then to compare them with the predictions of cosmological models, has been carried out by Hancock \ea (submitted) who find strong evidence for the
 existence of a Doppler peak on medium scales.
Further details together with a comparison with a fuller range of cosmological models will be presented in Rocha {\it et al} (in preparation). 
For the current purpose of discussion of the value of $n$, and comparison with the results obtained 
from Tenerife and COBE alone, we present here a version of the results for $n$ found in  
Hancock \ea (submitted).

The precise form
of the Doppler peak depends on the nature of the dark matter, and the
values of $\Omega_0$, $\Omega_b$ and $\ho$. Thus
in order to use the medium-scale anisotropy results to provide additional
leverage on the spectral index determination it is necessary to adopt
a given cosmological model. Taking the minimalist assumption of the
standard Cold Dark Matter (CDM) model, without gravity waves,
with $\ho=50 \kmspmpc$,
$\Omega_b=0.07$, Hancock \ea (submitted) find this offers a good
fit to all of the data. Tilting the initial spectrum of fluctuations
away from scale invariance and fitting to the $\Delta T_{l}$ we can delimit $n$ using the full data set.

Hancock \ea considered the four-year COBE data, and used the binned angular power spectrum (Tegmark 1996) in conjunction with Tenerife, Python, South-Pole, Saskatoon, MAX, ARGO, MSAM and CAT experiments.
As seen in Figure 8, only models with initial spectra in the range
$1.0 \le n \le 1.2$ are allowed by this improved analysis, with best fit values 
of $\qrms =15$ $\mu$K for $n=1.1$. These
results are consistent with those from large scales alone, and with
the inflationary value of $n\simeq 1.0$.
We note that the presence of a gravity wave component in our model
would require even larger values of $n$ than those derived above and
given Equation \ref{eq:tensors} this reaffirms our conclusion that
a significant gravity wave contribution is unlikely.
The above constraint on $n$ depends on which values of $\Omega_0$, $\Omega_b$
and $\ho$ are chosen, but is not simply related to any one parameter.
Constraints on $n$ derived using a fuller set of models are
given in Rocha {\it et al} (in preparation).

Prior to the COBE detection the most common normalization of models used was
 based  on the value of $\sigma _8$( $(\Delta M/M)_{rms}$
in a sphere of radius $8h^{-1}$Mpc)
derived from galaxy clustering assuming a bias $b=1$ \ie that light
traces mass (Kaiser 1984, Davis \ea 1985).
If in fact galaxies are more highly clustered than matter, the amplitude of the
initial matter perturbations (and hence \dtotrms) necessary to produce
the observed clustering is reduced by the factor $b$.
The Tenerife and
COBE data provide an independent and more accurate value for the normalisation
which for a given cosmological model
allows us to determine the degree of bias necessary for
consistency with the large scale structure observations.

In particular the COBE normalized CDM model with a tilt of $n=1.1$ gives a bias of the order $\simeq 0.7$.

\begin{figure}
\psfig{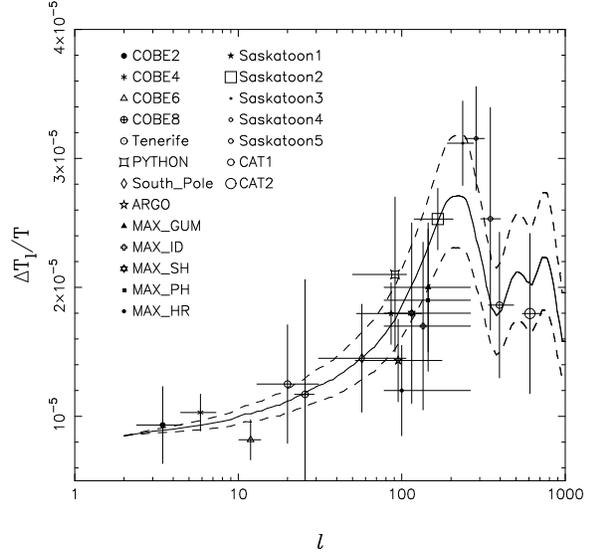}
\caption{Recent CMB anisotropy results on large and
intermediate angular scales are used to delimit the spectral index
$n$ of the primordial fluctuations.}
\label{Figure8}
\end{figure}

\section{Discussion and Conclusions}

We have presented a detailed analysis of the measurements taken at
Dec=+40\degg~by the Tenerife experiments at 10, 15 and 33 GHz.
After accounting for both local atmospheric and discrete radio source
foregrounds, the 15
and 33 GHz data at high Galactic latitude are seen to contain
statistically significant signals which have their origin in
common hot and cold features. The cross-correlation function between
the data at these two frequencies demonstrates that the amplitudes and
shapes of the structures detected at 15 and 33 GHz are similar. This,
combined with our measurements at the lower frequency of 10 GHz,
implies that the CMB
signal dominates over the Galactic contribution at 15 GHz, and that the
maximum possible Galactic contribution at 33 GHz is smaller than 10 \%
of the detected signal. 
Our best estimate of the cosmological signal is $\qrms=22^{+10}_{-6} \mu$K
for an $n=1$ inflationary spectrum.
This amplitude is reduced by $4\mu$K over that previously reported for the same
data set and results from an improved separation of signal from
atmospheric noise. 
Comparison of the Tenerife and COBE two year anisotropy detections
by means of the likelihood function allows a detailed investigation of
the allowed parameter space for a power law model of the fluctuation spectrum.
The best fit values of $n$ and \qrms are $1.37$ and $16 \mu$K,
and marginalising over
\qrms we find both data sets consistent with $n$ in the range $1.0
\le n \le 1.6$.
These results support inflationary models, which predict $n \simeq 1$
and future improvements in the Tenerife data and use of the 4 year COBE data will narrow
the range of allowed $n$. Improvements of this kind are important to determine
the power spectrum normalisation, since {\em only} on these large
angular scales is it possible to place limits on the tensor to scalar ratio
independent of the precise details of the cosmological model.
By combining
the large scale anisotropy measurements from Tenerife and COBE
with observations of medium scale anisotropy
measurements, improved limits have been placed on $n$, but at the
expense of assuming an underlying cosmological model (in our case CDM).

For a CDM model with $\ho=50 \kmspmpc$, $\Omega_{b}=0.07$, we find that this data set is consistent with $n$ in the range  $1.0 \leq n \leq 1.2$. The best fit values of $n$ and \qrms are 1.1 and $15 \mu$K. This COBE normalized tilted model predicts a bias of the order $\simeq 0.7$.
So although this model fits well the CMB data alone it may not give a realistic scenario when we consider jointly the CMB data and the observed galaxy clustering.
   
Our attempts to place increasingly more accurate limits on fundamental
cosmological parameters will undoubtedly place increasing demands on
observers for ever improved accuracy until the point where the intrinsic
cosmic variance becomes the dominant form of error. Confidence in the results
to this tolerance level will probably require close co-operation between
observers and the combination of results from space-based, balloon-based
and ground-based telescopes working over a range of frequencies.
Potentially the most powerful observations will result from mapping
overlapping/
interlocking CMB fields with independent multi-frequency instruments.
In contrast to the statistical results currently being reported,
this method builds in the necessary redundancy to reduce systematic
errors to an acceptable level. Such observations are currently in progress
with the Tenerife instruments, which in conjunction with the COBE 4 year data
should provide a high signal to noise
map of the last scattering surface thus providing a two dimensional
representation of the seed structures as compared to the simple
1-D scans reported here. Being on scales greater than
the horizon size at recombination the form of such a map would reflect
the structures generated in an inflationary driven phase in the very early universe.

\subsection*{ACKNOWLEDGMENTS}

\noindent The Tenerife experiments are supported by the UK Particle
Physics and Astronomy Research Council, the European Community Science
programme contract SCI-ST920830, the Human Capital and Mobility
contract CHRXCT920079 and the Spanish DGICYT science programme. S.
Hancock wishes to acknowledge a Research Fellowship at St. John's
College, Cambridge, G. Rocha wishes to acknowledge a JNICT 
studentship from Portugal.

\end{document}